\documentclass{article}

\usepackage{latexsym}
\usepackage{graphicx}
\usepackage{multicol,multirow}
\usepackage{amsmath,amssymb,amsfonts}
\usepackage{mathrsfs}
\usepackage{amsthm}
\usepackage{apacite}
\usepackage{rotating}
\usepackage{appendix}
\usepackage[authoryear]{natbib}
\usepackage{ifpdf}
\usepackage[T1]{fontenc}
\usepackage{times}
\usepackage{sourcesanspro}
\usepackage{newtxmath}
\usepackage{textcomp}%
\usepackage{xcolor}%
\usepackage{hyperref}
\usepackage[ruled,vlined]{algorithm2e}
\usepackage{algorithmic}
\usepackage{lineno}
\nolinenumbers

\title{Modelling and simulating spatial extremes by combining extreme value theory with generative adversarial networks}
\author{Younes Boulaguiem$^1$\footnote{Corresponding author: younes.boulaguiem@unige.ch.} \and Jakob Zscheischler$^{2,3,4}$ \and Edoardo Vignotto$^1$ Karin van der Wiel$^5$ Sebastian Engelke$^1$}
\date{%
    $^1$Geneva School of Economics and Management, University of Geneva, Boulevard du Pont d’Arve 40, Geneva, 1205, Switzerland\\%
    $^2$Department of Computational Hydrosystems, Helmholtz Centre for Environmental Research, Permoserstr. 15, 04318 Leipzig, Germany\\%
    $^3$Climate and Environmental Physics, University of Bern, Sidlerstrasse 5, 3012 Bern, Switzerland\\%
    $^4$Oschger Centre for Climate Change Research, University of Bern, 3012 Bern, Switzerland\\%
    $^5$Royal Netherlands Meteorological Institute, De Bilt, The Netherlands\\%
}

\begin{document}
\maketitle

\abstract{Modelling dependencies between climate extremes is important for climate risk assessment, for instance when allocating emergency management funds. In statistics, multivariate extreme value theory is often used to model spatial extremes. However, most commonly used approaches require strong assumptions and are either too simplistic or over-parameterized. From a machine learning perspective, Generative Adversarial Networks (GANs) are a powerful tool to model dependencies in high-dimensional spaces. Yet in the standard setting, GANs do not well represent dependencies in the extremes. Here we combine GANs with extreme value theory (evtGAN) to model spatial dependencies in summer maxima of temperature and winter maxima in precipitation over a large part of western Europe. We use data from a stationary 2000-year climate model simulation to validate the approach and explore its sensitivity to small sample sizes. Our results show that evtGAN outperforms classical GANs and standard statistical approaches to model spatial extremes. Already with about 50 years of data, which corresponds to commonly available climate records, we obtain reasonably good performance. In general, dependencies between temperature extremes are better captured than dependencies between precipitation extremes due to the high spatial coherence in temperature fields. Our approach can be applied to other climate variables and can be used to emulate climate models when running very long simulations to determine dependencies in the extremes is deemed infeasible. 
}

\section{Introduction}

Understanding and modelling climate extremes such as floods, heatwaves and heavy precipitation, is of paramount importance because they often lead to severe impacts on our socio-economic system \citep{IPCC2012}. Many impacts are associated with compounding drivers, for instance multiple hazards occurring at the same time at the same or different locations affecting the same system \citep{leonard2014compound,zscheischler2018future}. Ignoring potential dependencies between multiple hazards can lead to severe misspecification of the associated risk \citep{zscheischler2017dependence,Hillier2020}. However, estimating dependencies between extremes is a challenging task, requiring large amounts of data and/or suitable approaches to model  the phenomena of interest in a computationally feasible time.

A particularly challenging class of events are spatially compounding events \citep{Zscheischler2020}. Spatially compounding events occur when multiple connected locations are affected by the same or different hazards within a limited time window, thereby causing an impact. The compounding is established via a system capable of spatial integration, which accumulates hazard impacts in spatially distant locations. For instance, the global food systems are vulnerable to multiple co-occurring droughts and heatwaves in key crop-producing regions of the world \citep{Anderson2019,Mehrabi2019}. Similarly, spatially extensive floods \citep{Jongman2014} or sequences of cyclones \citep{Raymond2020} can deplete emergency response funds. Climate extremes can be correlated over very large distances due to teleconnections \citep{Boers2019} but modelling such dependencies is challenging.

One approach to tackle the challenge of spatially correlated extremes is to create a large amount of data by running very long simulations with state-of-the-art climate models, which have the physical spatial dependencies built into them. For instance, over the recent years for a number of climate models large ensembles have been created \citep{deser2020insights}. However, these simulations typically have rather coarse resolution, are usually not stationary in time and are very expensive to run. 

Extreme value theory provides mathematically justified models for the tail region of a multivariate distribution $(X_1,\dots, X_d)$, $d\geq 2$. This enables the extrapolation beyond the range of the data and the accurate estimation of the small probabilities of rare (multivariate) events. Statistical methods building on this theory are popular in a broad range of domains such as meteorology \citep{le2018dependence}, climate science \citep{naveau2020statistical} and finance \citep{poon2003extreme}. Applications have so far been limited to low-dimensional settings for several reasons.
On the one hand, even for moderately large dimensions $d$, the fitting and simulation of parametric models is computationally intensive because it requires computing complex likelihoods \citep{dom2016a}. On the other hand, the extremal dependence structures in applications are difficult to model and the existing approaches are either simplistic or over-parameterized. In spatial applications, for instance, Euclidean distance is used to parametrize the isotropic dependence of stationary max-stable random fields \citep{bla2011}. Most real world datasets that cover larger spatial domains do, however, feature non-stationarities in space that cannot be captured by the stationarity assumed in current geostatistical models; see \cite{eng2021} for a review on recent extreme value methods in higher dimensions.

Methods from machine learning are well-suited for complex and non stationary data sets. Their loss functions are, however, typically designed with the purpose to predict well in the bulk of the distribution. It is therefore a difficult problem to construct approaches with an accurate performance outside the range of the training data. In prediction problems, one possibility is to adapt the loss function to make the algorithm more sensitive to extreme values, as for instance done in quantile regression forests \citep{meinshausen2006quantile} or extreme gradient boosting \citep{velthoen2021gradient} for the prediction of high conditional quantiles. \citet{jalalzai2018binary} discuss this problem for classification in extreme regions and propose a new notion of asymptotic risk that helps to define classifiers with good generalization capacity beyond the observations' range in the predictor space.

Rather than predicting an unknown response, in this work we are interested in a generative model that is able to learn a high-dimensional distribution $(X_1,\dots, X_d)$ both in the bulk and in the extremes. We concentrate on Generative Adversarial Networks (GANs), which are known to be competitive models for multivariate density estimation. While classical applications of GANs are often in the field of image analysis \citep{zhang2017stackgan,karras2017progressive,choi2018stargan}, they have been used much more broadly in domains such as finance \citep{efimov2020using}, fraud detection \citep{zheng2019one}, speech recognition \citep{sahu2019modeling} and medicine \citep{schlegl2017unsupervised}.

There are two different aspects that characterize the extremal properties of multivariate data: the univariate tail behaviour of each margin and the extremal dependence between the largest observations.
For GANs, it has recently been shown \citep{wie2019, hus2021} that the marginal distributions of the generated samples are either bounded or light-tailed if the input noise is uniform or Gaussian, respectively. \citet{hus2021} propose to use heavy-tailed input to overcome this issue.
\cite{bhatia2020exgan} propose the ExGAN algorithm that uses conditional GANs to perform importance sampling of extreme scenarios. The main difference to our approach is that ExGAN simulates single extreme events, while our model will rely on block maxima and therefore does not suffer from serial correlation and seasonality issues. Moreover, ExGAN does not model the marginal distributions with extreme value theory and, since the input noise of ExGAN is Gaussian, this may be problematic especially for heavy-tailed data sets in view of the results of \cite{wie2019} and \cite{hus2021} mentioned above.

In this paper, we concentrate on modeling the extremal properties of both the marginal distributions and the dependence structure in a realistic way even for high-dimensional and spatially non-stationary data. Our model, called evtGAN, combines the asymptotic theory of extremes with the flexibility of GANs to overcome the limitations of classical statistical approaches to low dimensional and stationary data. By using extrapolation for the marginals from extreme value theory, our method can be applied to data sets with arbitrary combinations of light-tailed and heavy-tailed distributions. We use a stationary 2000-year climate simulation to test and validate our approach.

In sum, in this paper we demonstrate how our method combines the best of extreme value theory and generative adversarial networks in order to efficiently and accurately learn complex multidimensional distributions with non-trivial extremal dependence structures. Furthermore, one can easily sample from these learned distributions, which facilitates the study of complex data such as spatially distributed climate extremes with spatially highly heterogeneous dependence structure. The remainder of the article is structured as follows: in Section~\ref{sec:meth} we briefly lay out the theoretical background of extreme value theory and generative adversarial networks, and present the main algorithm of our methodology. Section~\ref{sec:appl} contains a description of the data and the model's architecture and hyper-parameters used for its training, and a presentation of the obtained results, which we then discuss in Section~\ref{sec:dis}. Finally, concluding remarks are laid out in Section~\ref{sec:conc}. 

\section{Methodology}
\label{sec:meth}
In the climate community there is a great need for methods able to efficiently provide an empirical description of the climate system, including extreme events, starting from as few ensemble runs as possible \citep{castruccio2019reproducing,deser2020insights}. Emulation techniques that tackle this challenge have been proposed in the last few years, usually focusing on the spatial correlation properties of the phenomena of interest \citep{mckinnon2017observational,mckinnon2018internal,link2019fldgen,Brunner2021}. In this work, our aim is to build an emulator specifically designed for extreme events that is able to reproduce the spatial tail dependencies of the data and extrapolate outside the range of the training samples. Our method does not require to discard any simulations and is therefore highly efficient. In this section we first recall some background on extreme value theory and generative adversarial networks, and then describe our algorithm called evtGAN. Similar to a copula approach \citep[e.g.][]{Brunner2021}, it relies on the idea to disentangle the modeling of the marginal distributions and the extremal dependence structure in order to use extreme value theory for the former and generative adversarial networks for the latter.

\subsection{Extreme value theory}
\label{sec:evt}

Extreme value theory provides mathematically justified methods to analyse the tail of a $d$-dimensional random vector $\mathbf X = (X_1,\dots, X_d)$. This branch of statistical theory has been widely applied in climate science, for example to study droughts \citep{burke2010extreme}, floods \citep{asadi2018optimal}, heatwaves \citep{tanarhte2015heat} and extreme rainfalls \citep{le2018dependence}.

For multivariate data, there are two different aspects that determine the quality of an extrapolation method: the univariate tail behaviour of each margin and the extremal dependence between the largest observations. We explain how to measure and model those based on componentwise maxima.
Let $\mathbf X_i = (X_{i1}, \dots, X_{id})$, $i=1,\dots, k$, be a sample of $\mathbf X$ and let
$\mathbf M_k = (M_{k1}, \dots, M_{kd})$, where $M_{kj} = \max(X_{1j}, \dots, X_{kj})$ is the maximum over the $j$th margin.

\subsubsection{Univariate theory}

The Fisher--Tippett--Gnedenko theorem \citep[e.g.,][]{coles2001introduction} describes the limit behaviour of the univariate maximum $M_{kj}$. It states that if there exist sequences $a_{kj}\in\mathbb R$, $b_{kj} > 0$ such that
\begin{align}\label{GEV_conv}
 \lim_{k\to\infty} \mathbb P\left(\frac{M_{kj}-a_{kj}}{b_{kj}} \le z\right) =   G_j(z), \qquad n\rightarrow \infty,
  \end{align}
then if $G_j$ is the distribution function of a non-degenerate random variable $Z_j$, it is in the class of generalized extreme value (GEV) distributions with 
$$ G_j(z)=\exp \left[ -\left\{ 1+\xi_j\left(\frac{z-\mu_j}{\sigma_j}\right)\right\}_+^{-1/\xi_j} \right], \quad z\in \mathbb R, $$
where $x_+$ denotes the positive part of a real number $x$, and $\xi_j \in \mathbb R$, $\mu_j \in \mathbb R$ and $\sigma_j>0$ are the shape, location and scale parameters, respectively. The shape parameter is the most important parameter since it indicates whether the $j$th margin is heavy-tailed ($\xi_j > 0$), light-tailed ($\xi =0$) or whether it has a finite uper end-point ($\xi<0$).

It can be shown that under mild conditions on the distribution of margin $X_j$, appropriate sequences exist for \eqref{GEV_conv} to hold. The above theorem therefore suggests to fit a generalized extreme value distribution $\widehat G_j$ to maxima taken over blocks of the same lengths, which is common practice when modelling yearly maxima. The shape, location and scale parameters can then be estimated in different ways, including moment-based estimators \cite{hosking1985algorithm}, Bayesian methods \citep{yoon2010full} and maximum likelihood estimation \citep{hosking1985estimation}. We use the latter in this work. This allows extrapolation in the direction of the $j$th marginal since the size of a $T$-year event can be approximated by the $(1-1/T)$-quantile of the distribution $\widehat G_j$, even if $T$ is larger than the length of the data record.

\subsubsection{Multivariate theory}\label{sec:mult}
For multivariate data, the correct extrapolation of the tail not only depends on correct models for the marginal extremes, but also on whether the dependence between marginally large values is well captured.
For two components $X_i$ and $X_j$ with limits $Z_i$ and $Z_j$ of the corresponding maxima in \eqref{GEV_conv}, this dependence is summarized by the extremal correlation $\chi_{ij}\in[0,1]$ \citep{sch2003}
\begin{equation*}
  \chi_{ij} = \lim_{q\to1} \mathbb P(F_i(X_i) >  q \mid F_j(X_j)  > q) \in [0,1].
\end{equation*}
The larger $\chi_{ij}$, the stronger the dependence in the extremes between the two components, that is, between $Z_i$ and $Z_j$. If
$\chi_{ij}>0$ we speak of asymptotic dependence, and otherwise of asymptotic independence. While the extremal correlation is a useful summary statistic, it does not reflect the complete extremal dependence structure between $X_i$ and $X_j$. Under asymptotic dependence, the latter can be characterized by the so-called spectral distribution \citep{deh1977}.
Let $\tilde X_j = -1/\log F_j(X_j)$, $j=1,\dots, d$, be the data normalized to standard Fr\'echet margins. The spectral distribution describes the extremal angle of $(X_i, X_j)$, given that the radius exceeds a high threshold, that is,
$$H(x) = \lim_{u\to \infty} \mathbb P( \tilde X_i / (\tilde X_i + \tilde X_j) \leq x   \mid  \tilde X_i + \tilde X_j > u ), \quad x \in [0,1].$$
Under strong extremal dependence, the spectral distribution centers around $1/2$; under weak extremal independence, it has mass close to the boundary points $0$ and $1$.

In multivariate extreme value theory, a popular way to ensure a correct extrapolation of tail dependence under asymptotic dependence is by modeling the joint distribution $\mathbf Z = (Z_1,\dots, Z_d)$ of the limits in \eqref{GEV_conv} as a max-stable distribution with multivariate distribution function
\begin{equation}\label{mevd}
  \mathbb P(G_1(Z_1)\leq z_1,\dots,G_d(Z_d)\leq z_d) = \exp\{-\ell(z_1,\dots,z_d)\}, \quad z_1,\dots,z_d \in [0,1],
\end{equation}
where $\ell$ is the so-called stable tail dependence function \citep[e.g.,][]{de2007extreme}. This is a copula approach in the sense that the right-hand side is independent of the original marginal distributions $G_j$ and only describes the dependence structure. In practice, the $G_j$ are replaced by estimates $\hat G_j$ that are fitted to the data first. The right-hand side of~\eqref{mevd} is also called an extreme value copula \citep[e.g.,][]{seg2010}.

%  In the bivariate case, the variable $(X_1,X_2)$ can also be represented in the polar representation \citep{dutfoy2014multivariate}, i.e.,
%  \begin{equation*}
%      (R,W)=(X_1 + X_2,\frac{X_1}{X_1+X_2}) \in [0,\mathbb R)\times[0,1].
% \end{equation*}
%  The first quantity $R$ gives a distribution of the distance between $(X_1,X_2)$ and the origin and it is called radial component, while the second quantity $W$ is called the angular component. Conditionally on $R$ being large, observations for which $W\approx 0$ or $W\approx 1$ corresponds to data points for which only one between $X_1$ and $X_2$ is extreme, while if $W\approx 1/2$ corresponds to data points for which both $X_1$ and $X_2$ are extremes. For perfectly independent variable, only the first situation will occur, while for perfectly dependent variable is the opposite. Non deterministic relations give results in between these two cases. The histogram of $W$ conditionally on $R$ being large is thus a useful tool for inspecting the overall dependence structure of $(X_1,X_2)$.

A natural extension of max-stable distributions to the spatial setting is given by max-stable processes $\{Z(t) : t \in \mathbb R^d\}$ \citep[e.g.,][]{davison2012statistical}, where the maxima $Z_i = Z(t_i)$ are observed at spatial locations $t_i$, $i=1,\dots, d$. This type of models have been widely applied in many different areas such as flood risk \citep{eng2014b}, heat waves \citep{eng2017a} and extreme precipitations \citep{buishand2008spatial}. 
A popular parametric model for a max-stable process $Z$ is the Brown--Resnick process \citep{kab2009, eng2014}. This model class is parameterized by a variogram function $\gamma$ on $\mathbb R^d$. In this case, each bivariate distribution is in the so-called H\"usler--Reiss family \citep{hue1989}, and the corresponding dependence parameter is the variogram function evaluated at the distance between the two locations,
$$\lambda_{ij} = \gamma( \| t_i - t_j \|),\qquad i,j =1,\dots, d.$$ 
One can show that the extremal correlation coefficient for this model is given by $\chi_{ij} = 2 - 2\Phi(\sqrt{\lambda_{ij}} /2)$, where $\Phi$ is the standard normal distribution function. A popular parametric family is the class of fractal variograms that can be written as  $\gamma_{\alpha, s}(h) = h^\alpha/s$, $\alpha \in (0,2]$, $s>0$.
While these methods enjoy many interesting properties, they are often quite complex to fit \citep{dom2016a} and impose strong assumptions on the spatial phenomena of interest, such as spatial stationarity, isotropic behaviour and asymptotic dependence between all pairs of locations.

A simple way of fitting a Brown--Resnick process to data is to compute the empirical versions $\hat \chi_{ij}$ of the extremal correlation coefficients and then numerically find the parameter values (e.g., $\alpha$ and $s$ for the fractal variogram family) whose implied model extremal coefficients minimise the squared error compared to the empirical ones.

\subsection{Generative adversarial network}
\label{sec:GANs}
Generative adversarial networks \citep{goodfellow2014generative} are attracting great interest thanks to their ability of learning complex multivariate distributions that are possibly supported on lower-dimensional manifolds. These methods are generative since they allow the generation of new realistic samples from the learnt distribution that can be very different from the training samples. 

Given observations $\mathbf U_1, \dots, \mathbf U_n$ from some $d$-dimensional unknown target distribution $p_{data}$, 
generative adversarial networks can be seen as a game opposing two agents, a discriminator $D$ and a generator $G$. 
The generator $G$ takes as input a random vector $\mathbf Y= (Y_1,\dots, Y_p)$ from a $p$-dimensional latent space and transforms it to a new sample $\mathbf U^* = G(\mathbf Y)$. The components $Y_j$, $j=1,\dots, p$, are independent and typically have a standard uniform or Gaussian distribution.

The discriminator $D$ has to decide whether a new sample $\mathbf U^*$ generated by $G$ is fake and following the distribution $p_G$, or coming from the real observations with distribution $p_{data}$. The discriminator expresses its guesses with a value between 0 and 1 corresponding to its predicted probability of the sample coming from the real data distribution $p_{data}$. Both the generator and the discriminator become better during the game, in which $D$ is trained to maximise the probability of correctly labelling samples from both sources, and $G$ is trained to minimise $D$’s performance and thus learns to generate more realistic samples.

Mathematically, the optimization problem is a two-player minimax game with cross-entropy objective function %\citep[see,][]{goodfellow2014generative}:
\begin{equation}
\min\limits_{G}\max\limits_{D} \mathbb{E}_{\mathbf U \sim p_{data}} [\log(D(\mathbf U))] + \mathbb{E}_{\mathbf Y\sim p_{\mathbf Y}} [\log(1-D(G(\mathbf Y))], \label{eq:goodfel}
\end{equation}
where $\mathbf Y$ is a random vector sampled from a latent space. In equilibrium, the optimal generator satisfies $p_G = p_{data}$ \citep{goodfellow2014generative}.

In practice, the discriminator and the generator are modeled through feed-forward neural networks. While the equilibrium with $p_G = p_{data}$ is only guaranteed thanks to the convex-concave property in the non-parametric formulation~\eqref{eq:goodfel}, the results remain very good as long as suitable adjustments to the losses, training algorithm and overall architecture are made to improve on the stability and convergence of training. For instance, a standard adjustment to the generator's loss function was already proposed in the GANs original paper \citep{goodfellow2014generative}, and consists on training the generator to minimise $ -\log\left(D\left(G\left(\mathbf Y\right)\right)\right)$ rather than $\log\left(1 - D\left(G\left(\mathbf Y\right)\right)\right)$. This new loss function is called the non-saturating heuristic loss, and was suggested to mitigate close to null gradients that occur in early training. Indeed, as the generator is still poorly trained, the discriminator can easily detect its samples and therefore associates values close to zero to them. This results in a very slow improvement of the generator during backpropagation as the gradients associated with values around $D\left(G\left(\mathbf Y\right)\right) = 0$ are too small, and thus its loss $L\left(G\right)=\log\left(1 - D\left(G\left(\mathbf Y\right)\right)\right)$ saturates. Training the generator to minimise $-\log\left(D\left(G\left(\mathbf Y\right)\right)\right)$ rather than $\log\left(1 - D\left(G\left(\mathbf Y\right)\right)\right)$ provides stronger gradients in early training without compromising on the dynamics of the two agents. 

The architecture of the generator $G$ and the discriminator $D$ can be specifically designed for the structure of the data under consideration. In this direction, \citet{radford2015unsupervised} introduces the use of the convolutional neural network as an add-on feature to the GAN yielding a deep convolutional generative adversarial network model (DCGAN). It is considered the preferred standard for dealing with image data and more generally any high dimensional vector of observations describing a complex underlying dependence structure, showing superior performance in the representation and recognition fields. 

\subsection{evtGAN}

Here we give an overview of evtGAN. Let $\mathbf X_1, \dots, \mathbf X_m$ be independent observations of dimensionality $d$, where $\mathbf X_i=\left(X_{i1},\ldots,X_{id}\right)$ and $m=k \cdot n$. Now let $\mathbf Z_1, \dots, \mathbf Z_n$ be independent observations obtained as block maxima taken sequentially over the span of $k$ observations, i.e., $\mathbf Z_1 = \max(\mathbf X_{1}, \dots, \mathbf X_{k})$, $\mathbf Z_2 = \max(\mathbf X_{k+1}, \dots, \mathbf X_{2k})$ and so on (see also Section~\ref{sec:evt}), where the maxima are taken in each component.
The evtGAN algorithm (Algorithm~\ref{alg:evt}) below takes these empirical block maxima as input and follows a copula approach where marginal distributions and the dependence structure are treated separately. In particular, this allows us to impose different amounts of parametric assumptions on the margins and the dependence.

It is known that classical GANs that are trained with bounded or light tailed noise input distributions in the latent space will also generate bounded or light-tailed samples, respectively \citep[see][]{wie2019, hus2021}. That means that they are not well-suited for extrapolating in a realistic way outside the range of the data. For this reason, we suppose that the approximation of the margins by generalized extreme value distributions $G_j$ as in \eqref{GEV_conv} holds, which allows for extrapolation beyond the data range.
On the other hand, we do not make explicit use of the assumption of a multivariate max-stable distribution as in \eqref{mevd}. This has two reasons.  First, while the univariate limit \eqref{GEV_conv} holds under very weak assumptions, the multivariate max-stability as in \eqref{mevd} requires much stronger assumptions \citep[see][]{res2008} and may be too restrictive for cases with asymptotic independence \citep[e.g.,][]{wadsworth2012dependence}. Second, even if the data follow a multivariate max-stable distribution \eqref{mevd}, the probabilistic structure is difficult to impose on a flexible machine learning model such as a GAN.

\begin{algorithm}
\caption{evtGAN}
\textbf{Input:} Observations $\mathbf Z_i = (Z_{i1}, \dots, Z_{id})$, $i=1,\dots, n$.
\begin{algorithmic}[1]
    \STATE For $j=1,\dots, d$, fit a GEV distribution $\widehat G_j$ to the data $Z_{1j},\dots, Z_{nj}$ with estimated parameters $(\hat \mu_j, \hat \sigma_j, \hat \xi_j)$.
    \STATE Normalize all margins empirically to a standard uniform distribution to obtain pseudo observations
    $$ \mathbf{U}_i = (\widehat F_1(Z_{i1}), \dots, \widehat F_d(Z_{id})), \quad i=1,\dots, n,$$
    where $\widehat F_j$ is the empirical distribution function of the $Z_{1j},\dots, Z_{nj}$.
    \STATE Train a DCGAN $G$ on the normalized data $\mathbf{U}_1,\dots, \mathbf{U}_n$ based on the loss in equation \eqref{eq:goodfel}.
    \STATE Generate $n^*$ new data points $\mathbf{U}_1^*,\dots, \mathbf{U}_n^*$ from $G$ with uniform margins, $i=1,\dots, n^*$.
    \STATE Normalize back to the scale of the original observations
    $$\mathbf{Z}_i^* = (\widehat G_1^{-1}(U_{i1}^*), \dots, \widehat G_d^{-1}(U_{id}^*)),\quad i=1,\dots, n^*.$$    
\end{algorithmic}
\textbf{Output:} Set of new generated observations $\mathbf Z^*_i = (Z_{i1}, \dots, Z^*_{id})$, $i=1,\dots, n^*$.
\label{alg:evt}
\end{algorithm}

The margins are first fitted by a parametric GEV distribution (line 1 in Algorithm~\ref{alg:evt}). They are then normalized to (approximate) pseudo-observations by applying the empirical distributions functions to each margin (line 2). This standardization to uniform margins stabilizes the fitting of the GAN. Alternatively, we could use the fitted GEV distributions for normalization but this seems to give slightly worse results in practice.
The pseudo-observations contain the extremal dependence structure of the original observations. Since this dependence structure is very complex in general and our focus is to reproduce gridded spatial fields, we do not rely on restrictive parametric assumptions but rather learn it by a highly flexible DCGAN (line 3). From the fitted model we can efficiently simulate any number of new pseudo-observations that have the correct extremal dependence (line 4). Finally, we transform the generated pseudo-observations back to the original scale to have realistic properties of the new samples also in terms of the margins (line 5).

\section{Application}
\label{sec:appl}

We assess the ability of the proposed method to correctly model the spatial tail dependence structure between climate extremes. Because observational records are relatively short and contain temporal trends, here we rely on 2000 years of climate model output that is representative of present-day weather over western Europe. In particular, we apply our approach to summer temperature and winter precipitation maxima, yielding 2000 and 1999 maxima for temperature and precipitations, respectively. We then compare the performance of our algorithm to the Brown--Resnick model, a state-of-the-art statistical model for spatial extremes. We use the fractal variogram family and the fitting described in Section~\ref{sec:mult}.
In order to retrieve a non-biased estimate of the error that each of the method makes, we divide the data into a training set of 50 observations, and the rest is taken as a test set. Both methods are fitted using only the training data set and the evaluation is made considering the information contained in the test set as ground truth. When not stated otherwise, the results from the Brown--Resnick model are analytical, while the results from evtGAN are obtained simulating $10'000$ data points.% and computing empirically the quantities of interest from these realizations.

\subsection{Data}
The model experiment, which uses large ensemble simulations with the EC-Earth global climate model \citep[v2.3,][]{hazeleger2012} has been widely used in climate impact studies \citep{van2019meteorological,kempen2021,Tschumi2021,Vogel2021,vanderwiel2021} and was originally designed to investigate the influence of natural variability on climate extremes. A detailed description of these climate simulations is provided in \citet{vanderWiel2019}, here we only provide a short overview. 
The large ensemble contains 2000~years of daily weather data, representative of the present-day climate. Present-day was defined by the observed value of global mean surface temperature (GMST) over the period 2011-2015 \citep[HadCRUT4 data,][]{morice2012}; the five-year time slice in long transient model simulations (forced by historical and Representative Concentration Pathway (RCP) 8.5 scenarios) with the same GMST was simulated repeatedly. To create the large ensemble, at the start of the time slice, twenty-five ensemble members were branched off from the sixteen transient runs. Each ensemble member was integrated for the five-year time slice period. Differences between ensemble members were forced by choosing different seeds in the atmospheric stochastic perturbations \citep{buizza1999}. This resulted in a total of $16 \times 25 \times 5 = 2000$~years of meteorological data, at T159 horizontal resolution (approximately 1$^\circ$), among which we selected temperature (Kelvin) and precipitations (meter/day) for this paper. 

We choose an area such that it is big enough to be relevant for climate application while being small enough to ensure a fair comparison of our approach and the classical approach in statistics to model spatial tail dependence, the Brown--Resnick process. Our analysis thus focuses on a total of $18\times 22$ grid points covering most of western Europe. 
For that area we compute for each grid point the summer temperature maxima and winter precipitation maxima.

\subsection{Architecture and hyper-parameters}
\label{sec:arch}

In our application, we consider the non-saturating heuristic loss \citep{goodfellow2014generative} for the generator, and the standard empirical cross-entropy loss for the discriminator. 
Training is done iteratively, where the discriminator is allowed to train two times longer than the generator. A bigger number of discriminator training iterations per generator training iteration was initially considered but it did not achieve better results for the considered sample sizes and input dimensions but only resulted in longer training.
We incorporate an exponential moving average scheme \citep{gidel2018variational} to the training algorithm as it has demonstrated more stability, a far better convergence and improved results.

\begin{figure}[h]
    \centering
        \includegraphics[width=.91\textwidth]{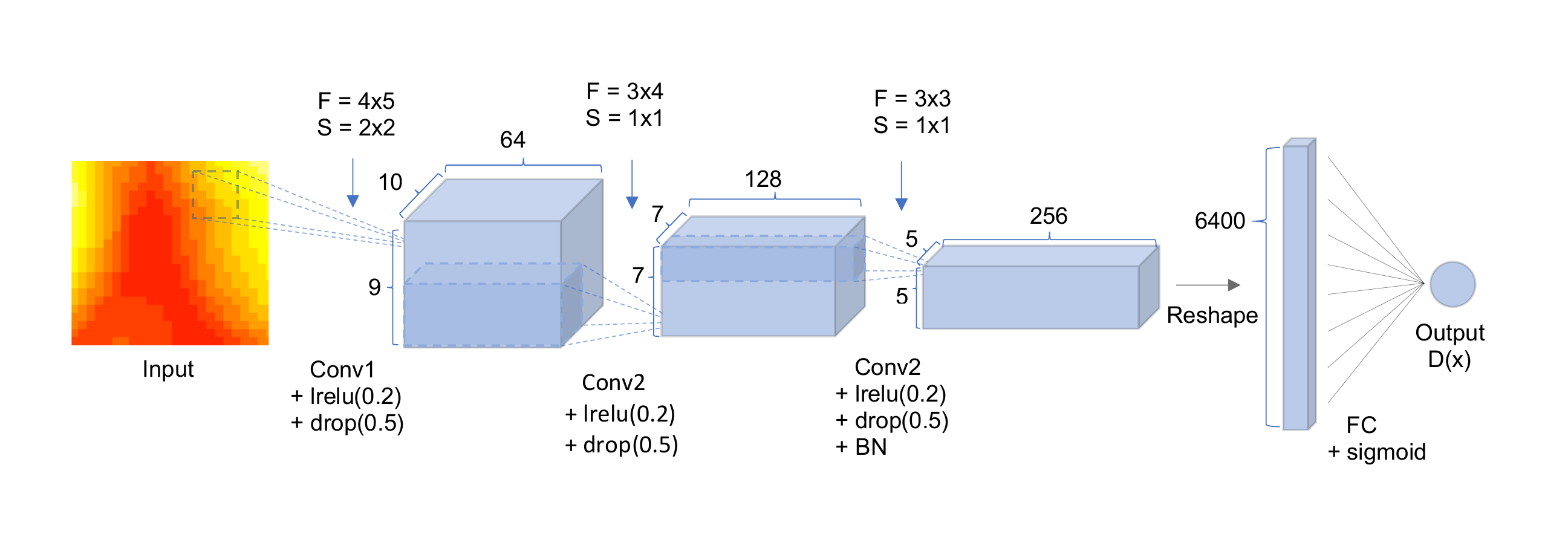}
        \caption{Discriminator's architecture and values of hyper-parameters used for training. \textit{lrelu} stands for the leaky relu function and its argument corresponds to the value of the slope when $x<0$; \textit{drop} stands for drop-out (see Section~\ref{sec:arch}) and its argument corresponds to the drop-out probability; \textit{F} stands for the kernel's dimension, and \textit{S} for the value of the stride; \textit{FC} stands for fully connected, and \textit{sigmoid} for the sigmoid activation function}
    \label{fig:discrim}
\end{figure}

\begin{figure}[h]
    \centering
        \includegraphics[width=.91\textwidth]{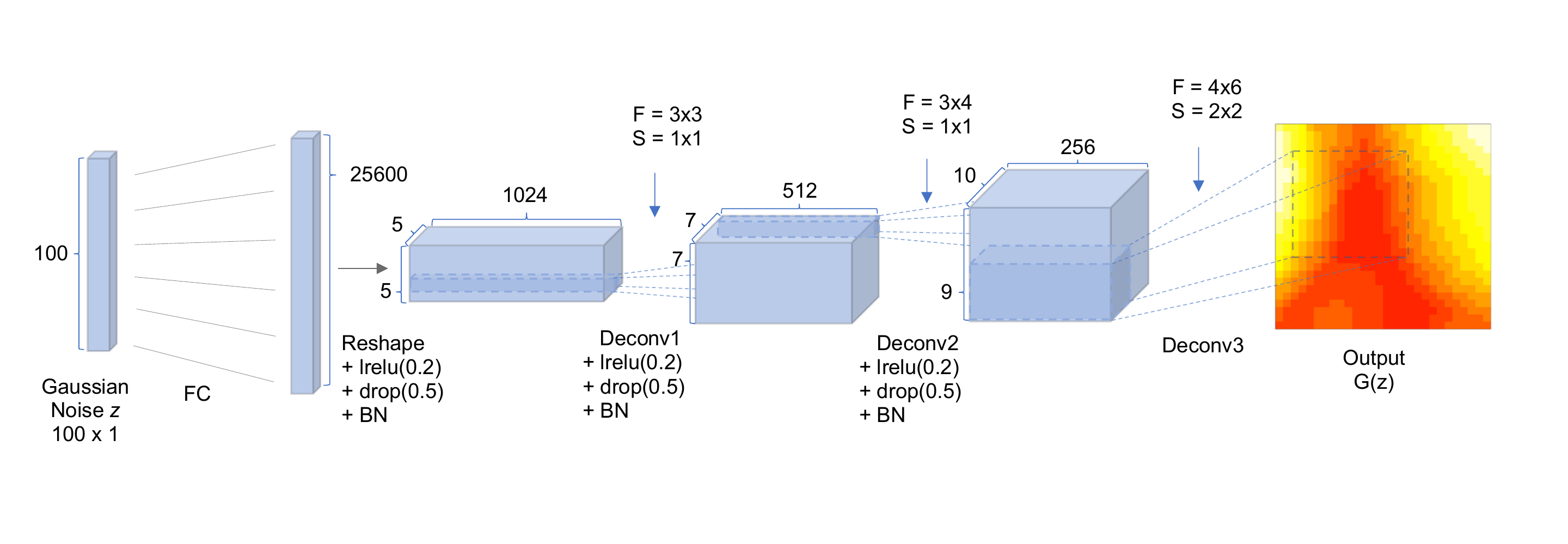}
        \caption{Generator's architecture and values of hyper-parameters used for training. \textit{lrelu} stands for the leaky \textit{relu} function and its argument corresponds to the value of the slope when $x<0$; \textit{drop} stands for drop-out (see Section~\ref{sec:arch}) and its argument corresponds to the drop-out probability; \textit{F} stands for the kernel's dimension, and \textit{S} for the value of the stride; \textit{BN} stands for batch-normalisation (see Section~\ref{sec:arch}); \textit{FC} stands for fully connected}
    \label{fig:gener}
\end{figure}

Since we focus on reproducing gridded spatial fields, we make use of the DCGAN model \citep{radford2015unsupervised} as an architectural design to take advantage of convolution \citep{fukushima1980neocognitron}, and for a more accurate analysis of the input data, we apply zero-padding of a single layer to the train set observations before they are fed to the discriminator \citep{albawi2017understanding}, increasing the dimension of its input to $20\times24$.
We use classical regularisation tools such as drop-out \citep{srivastava2014dropout}, batch-normalisation \citep{ioffe2015batch} and the Adam optimizer \citep{kingma2014adam} with a learning rate of $2\times10^{-4}$ and batch size of $50$ to train the neural network, and a total of $30,000$ training epochs. %For each epoch the discriminator is updated twice more than the generator. 
The value of the exponential moving average parameter in the training algorithm is set to $\alpha=0.9$. The values of the hyper-parameters are the result of an extensive heuristic tuning, and are as such only valid for the dimension of the input and the considered sample sizes involved in training the network. That being said, we are fairly confident that a minimal intervention in tuning the hyper-parameters/architecture would be required if for instance higher resolution inputs were to be used, as these models are fairly robust to such changes. Indeed, models based on the general GANs paradigm are traditionally trained on much higher dimensional objects such as images, and they learn well using an architecture that is very close to the one we have adopted. One may argue that the underlying distributions defining real world images are "simpler", and therefore easier to learn, which is why we propose the evtGAN model to address the increased complexity of the underlying distribution of the data at hand: initially, we started with a coarser grid (lower resolution input), as we considered France only, then decided to cover a larger part of Europe as the results were very good. In this change, the model’s architecture needed to be adjusted only by adding a few layers to the generator. We therefore remain confident that a higher resolution grid would only entail a rather small hyper-parameter/architectural adjustment for the results to remain satisfactory. 
For a given input resolution, increasing the complexity of the model (using a deeper one) just increased training time and pushed convergence further down the line.
Figures \ref{fig:discrim} and \ref{fig:gener} illustrate the final architectures of the discriminator and the generator, as well as the values of some of the hyper-parameters used for training.
%Finally, for stability purposes, the training data are empirically and uniformly standardized in $(0,1)$.

\subsection{Results}
\begin{figure}
    \centering
        \includegraphics[width=.81\textwidth]{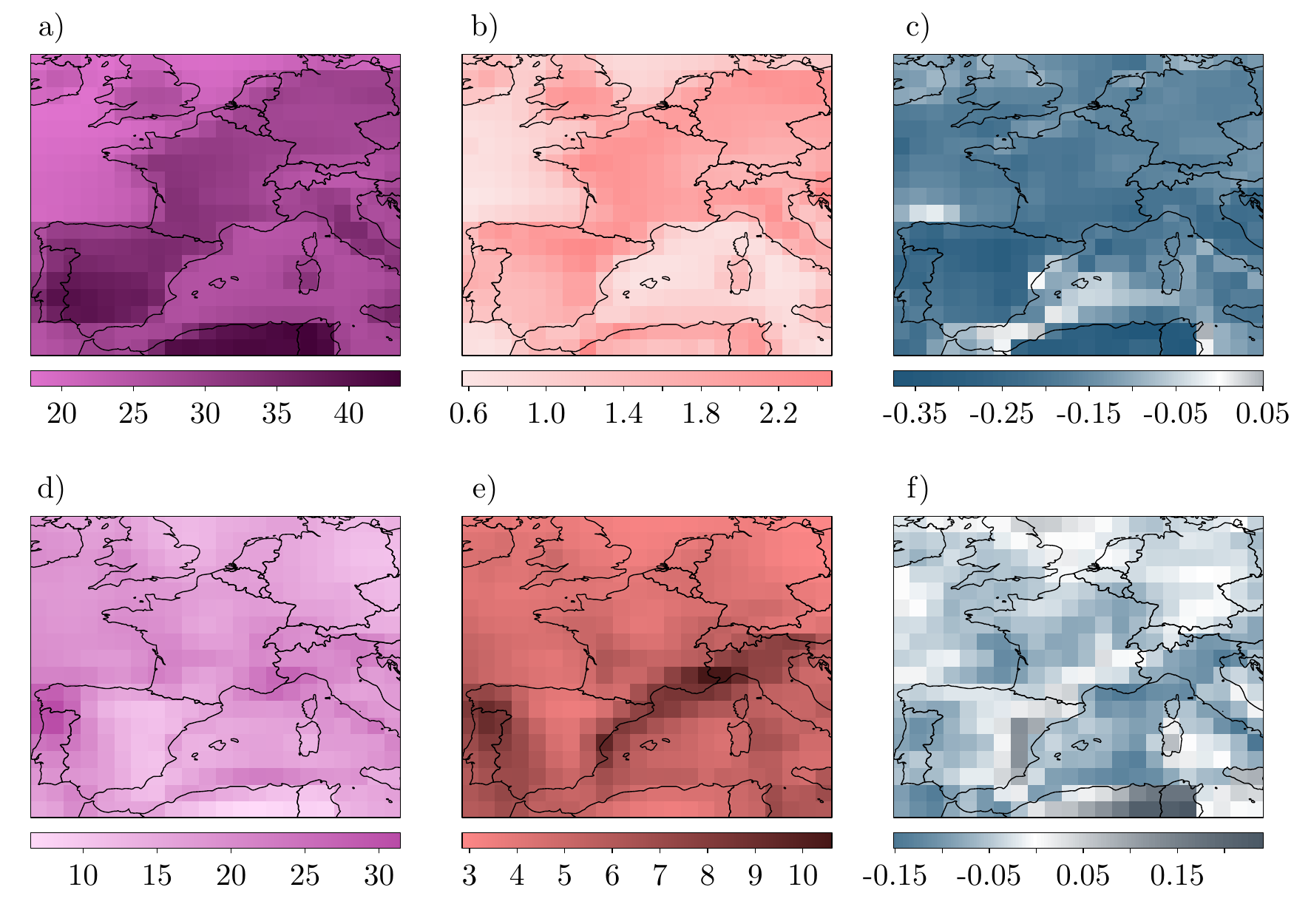}
        \caption{The generalized extreme value distribution parameters estimated for each grid point for temperature (a-c) and precipitation (d-f) extremes. On the left (a, d) the mean parameter $\mu$, in the middle (b, e) the scale parameter $\sigma$ and on the right (c, f) the shape parameter $\xi$}
    \label{fig:grid}
\end{figure}

A particularity of evtGAN is that it decomposes the full high-dimensional distribution of the data into its marginal distributions and its dependence structure and processes them separately. We thus first report estimated key parameters of the marginal distributions. 
For temperature, the location parameter $\mu$ is typically higher over land than over oceans (Fig.~\ref{fig:grid}a). Furthermore, there is a trend towards lower values of $\mu$ for more northern regions, illustrating the well-known latitudinal temperature gradient from south to north. The scale parameter $\sigma$, representing the width of the distribution, is also higher over land than over the ocean, with a fairly homogeneous distribution over land (Fig.~\ref{fig:grid}b). The shape parameter $\xi$ is well below 0 for most regions except some areas in the Atlantic north of Spain and in the Mediterranean at the north African coast, indicating a bounded tail (Fig.~\ref{fig:grid}c).

For precipitation, the location parameter $\mu$ is more similar between land and ocean compared to temperature but is a little bit more heterogeneous in space (Fig.~\ref{fig:grid}d), illustrating orographic effects on precipitation extremes. For the scale parameter $\sigma$ there is also no clear land-sea contrast but areas with relatively much higher values (corresponding to higher interannual variability in extremes) occur over the Western part of the Iberian peninsula and along a coastal band at the northern Mediterranean coast (Fig.~\ref{fig:grid}e). The shape parameter $\xi$ is spatially quite heterogeneous, with values ranging from -0.15 to 0.25 (Fig.~\ref{fig:grid}f). Overall, the parameter of the extreme value distributions of precipitation extremes show much higher spatial heterogeneity than the one for temperature extremes, suggesting that it might be more difficult to learn a model that represents well all tail dependencies for the entire region.

%Lastly, we evaluate the performance of the  marginal model comparing the average absolute errors of the retrieved return levels for different return periods with the corresponding ground truth value, i.e.
%\begin{equation*}
%    \sum_{i=1}^{396}\sum_{j=400}^{600} \mid \hat r_{ij} - r_{ij}\mid,
%\end{equation*}
%where $\hat r_{ij}$ is the estimated return level for a return period $j$ for the grid point $i$ and $r_{ij}$ is the ground truth. The average loss amount to $0.436$ years with a standard deviation of $0.010$ years, indicating that the considered training length (300 years) is enough to fully characterize the univariate information at each grid point, thus guarantying to our approach the ability to simulate extreme events that lie outside the training space.

\begin{figure}[h]
    \includegraphics[width=\textwidth]{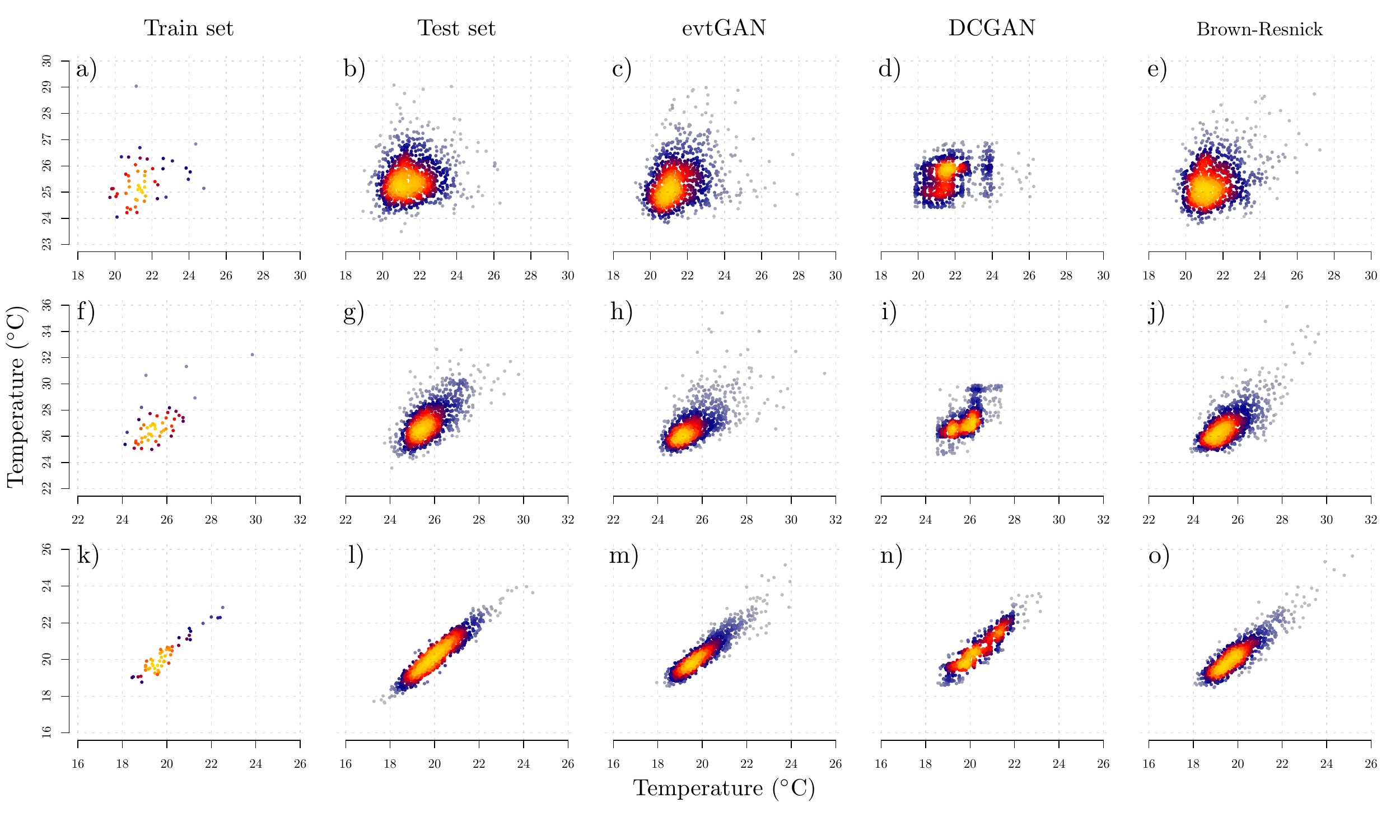}
    \caption{Bivariate plots of temperature extremes based on train and test sets and simulated through the different presented methods. Shown are selected pairs of locations with varying tail dependence. Columns from left to right: train, test, evtGAN, DCGAN and Brown--Resnick. From top to bottom: weak tail dependence (a-e), mild tail dependence (f-j), strong tail dependence (k-o). The colours represent the empirically estimated density: the spectrum goes from yellow to grey, reflecting a decrease in density}
    \label{fig:bivtemp}
\end{figure}

Next we look at examples of bivariate scatterplots of simulated temperature extremes from the different approaches for three cases with varying tail dependence (Fig.~\ref{fig:bivtemp}). The three rows correspond to three pairs of grid points with weak, mild and strong tail dependence, respectively. The train sets (Fig.~\ref{fig:bivtemp}a, f, k) illustrate the distributions from which the models are tasked to learn while the test sets (Fig.~\ref{fig:bivtemp}b, g, l) are used as ground truth. As can be clearly seen, extremes are much less likely to co-occur between locations that are characterized by weak tail dependence (Fig.~\ref{fig:bivtemp}b) compared to locations that are characterized by strong tail dependence (Fig.~\ref{fig:bivtemp}l). Purely based on visual judgement, it seems that evtGAN is able to characterize the different tail dependencies relatively well and can simulate samples outside of the range of the train set (Fig.~\ref{fig:bivtemp}c, h, m) whereas DCGAN only simulates distributions bounded to the range of the train set (Fig.~\ref{fig:bivtemp}d, i, n) and Brown--Resnick tends to overestimate tail dependence in cases where tail dependence is weak (Fig.~\ref{fig:bivtemp}e, j).

\begin{figure}[h]
    \includegraphics[width=\textwidth]{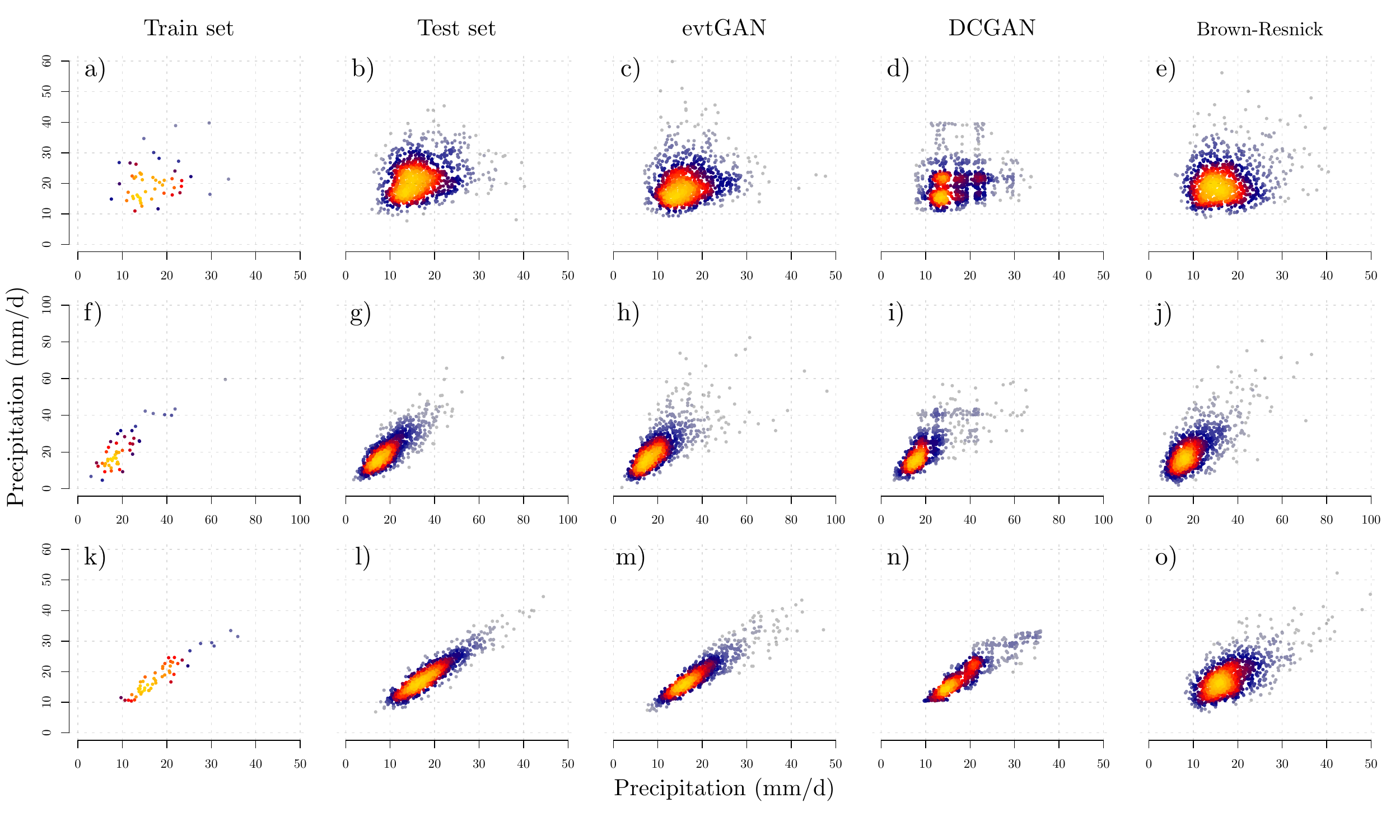}
    \caption{Bivariate plots of precipitation extremes based on train and test sets and simulated through the different presented methods. Shown are selected pairs of locations with varying tail dependence. Columns from left to right: train, test, evtGAN, DCGAN and Brown--Resnick. From top to bottom: weak tail dependence (a-e), mild tail dependence (f-j), strong tail dependence (k-o). The colours represent the empirically estimated density: the spectrum goes from yellow to grey, reflecting a decrease in density}
    \label{fig:bivrain}
\end{figure}

The corresponding figure for precipitation extremes is shown in Fig.~\ref{fig:bivrain}. Conclusions are similar as for temperature extremes. The evtGAN simulates distributions with different tail dependencies and samples that are far outside the range of the train set (Fig.~\ref{fig:bivrain}c, h, m). DCGAN simulates distributions that are bounded to the train set range (Fig.~\ref{fig:bivrain}d, i, n). On the other hand, Brown--Resnick overestimates tail dependence in the case of mild tail dependence (Fig.~\ref{fig:bivrain}j), and underestimates tail dependence when tail dependence is strong (Fig.~\ref{fig:bivrain}o).

\begin{figure}[h]
  \centering
  \includegraphics[width=\linewidth]{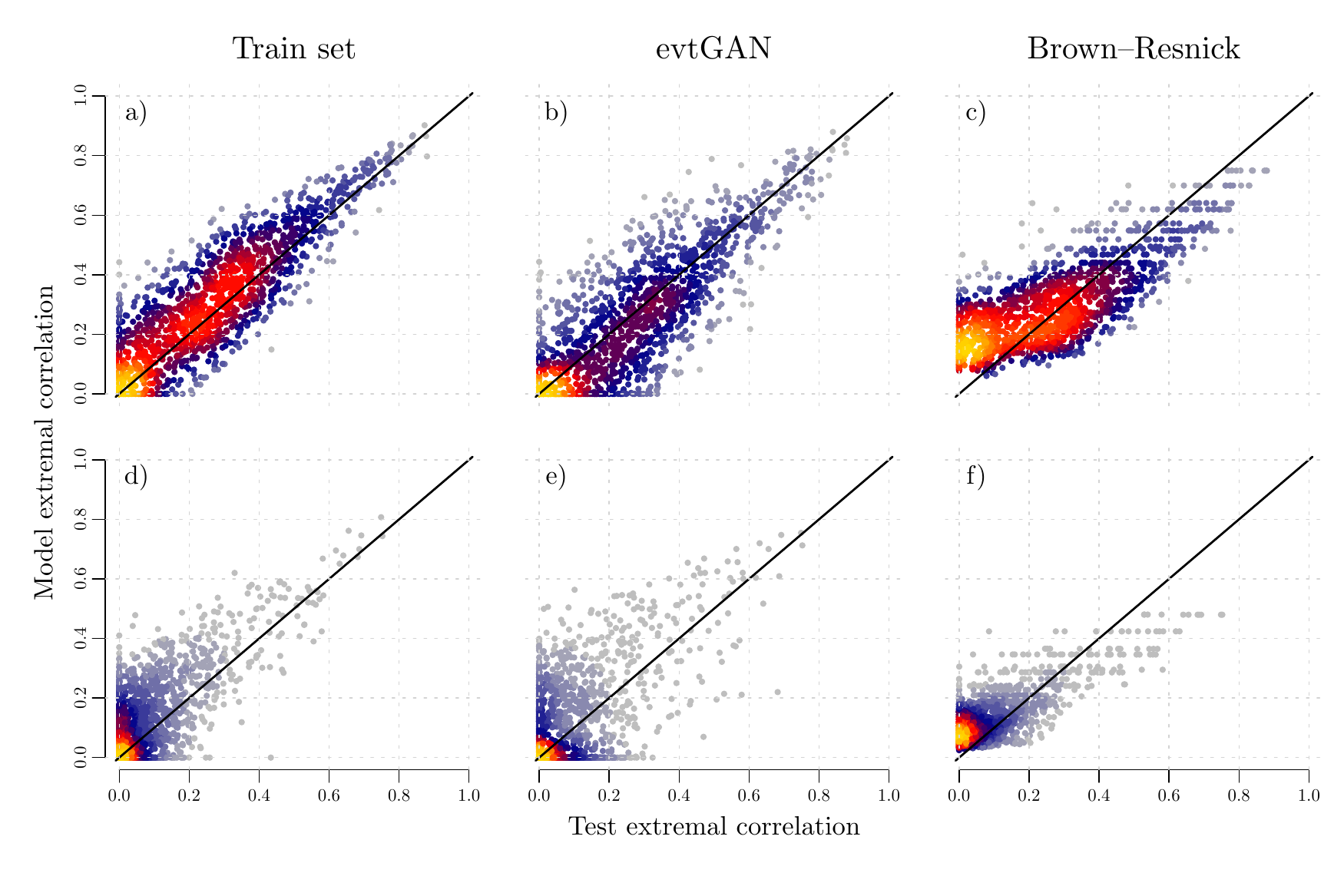}
    \caption{Scatter plots of the extremal correlations for temperature extremes (a-c) and precipitation extremes (d-f) between $100$ randomly selected locations. The x-axes always show the estimates based on the test set. Estimates on the y-axes are based on the train set (a, d), on the output of evtGAN (b, d) and on the Brown--Resnick model}
\label{fig:eval}
\end{figure}

A scatterplot of bivariate extremal correlations between 100 randomly selected locations estimated from the train set, evtGAN and Brown--Resnick, respectively, against estimates based on the test set (1950 samples) is shown in Fig.~\ref{fig:eval}. The estimates derived directly from the train sets (Fig.~\ref{fig:eval}a, d) are the benchmark, and by design better performance is not possible. Clearly, pairs of locations with stronger tail dependence are much more likely for temperature (Fig.~\ref{fig:eval}a) than for precipitation (Fig.~\ref{fig:eval}d), confirming the impression obtained from the spatial homogeneity of the parameters in the extreme value distributions (Fig.~\ref{fig:grid}). Furthermore, evtGAN seems to better capture the observed relationship. Brown--Resnick has difficulties in particular with pairs that have weak or no tail dependence (extremal correlation equals 0, lower left in the figures), which is consistent with Fig.~\ref{fig:bivtemp} and Fig.~\ref{fig:bivrain}. 

%We then evaluate the performance of the evtGAN in capturing the tail dependence structure of the data and compare it with a more common approach, i.e., the Brown-Resnick model. In Figure \ref{fig:eval} the scatter plots between the ground truth extremal coefficients between each pair of 100 randomly selected locations and the associated estimates for the two methods are compared. As it is visible, the evtGAN makes a low error for both strongly dependent and weakly pair, with no particular bias. As it is expectable, higher losses are associated with weak dependence, i.e., the more distant pair. On the contrary, the Brown--Resnick model makes, in general, higher error for every level of dependence. Furthermore, there is in this case a relevant bias for weakly dependence pairs. The average error for the evtGAN is $0.059$, while for the Brown-Resnick model it is $0.11$.

\begin{figure}[h]
\centering
  \includegraphics[width=0.91\linewidth]{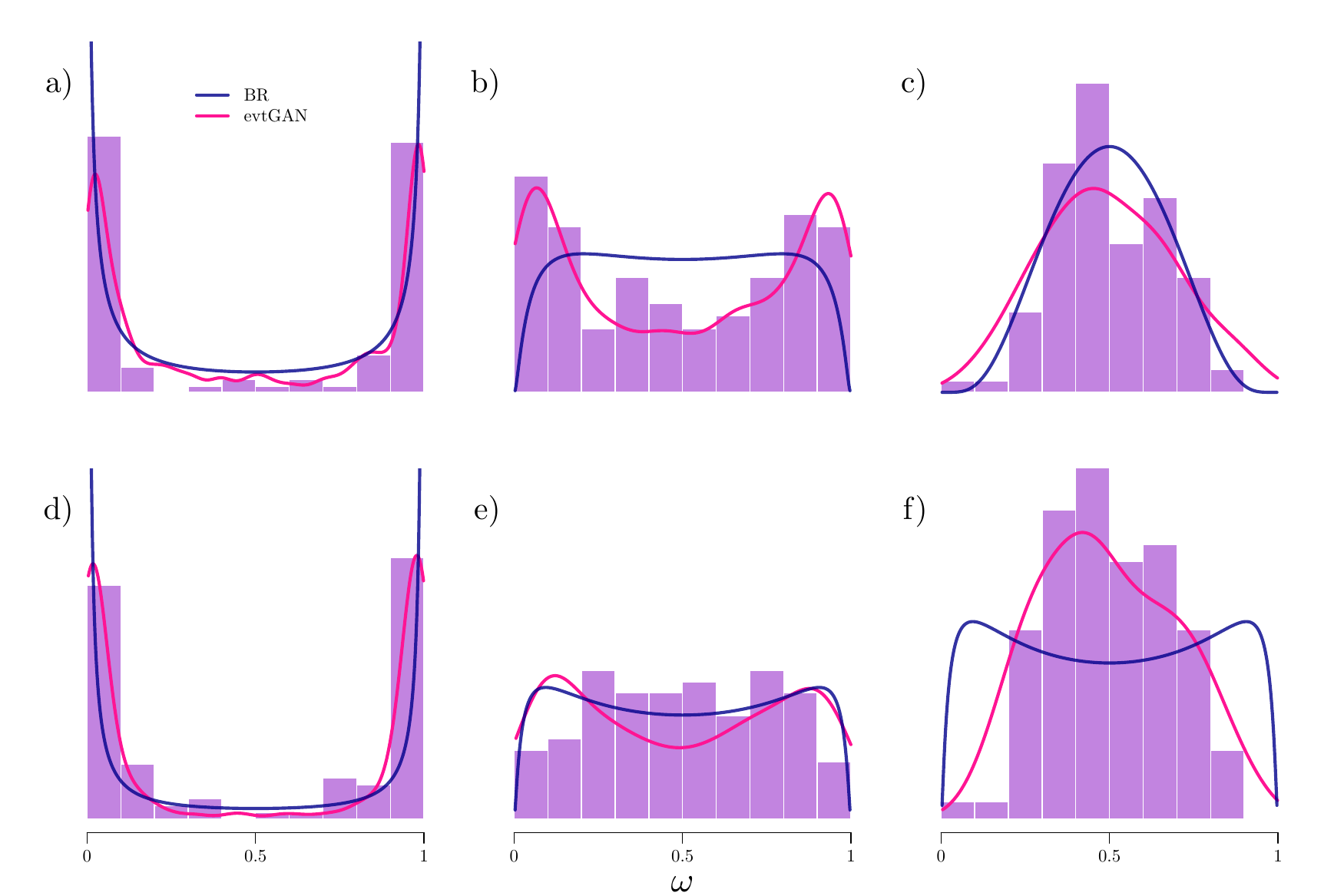}
\caption{Spectral distributions for a threshold of $0.95$ for selected pairs of locations with varying tail dependence for temperature (a-c) and precipitation (d-f). (a, d) weak tail dependence, (b, e) mild tail dependence, (c, f) strong tail dependence. In red the kernel density estimate of the evtGAN, in blue the Brown--Resnick model and in magenta bars the ground truth}
\label{fig:spectral}
\end{figure}

To further explore the behaviour of evtGAN and Brown--Resnick in dealing with different levels of dependence, Fig.~\ref{fig:spectral} shows a comparison of the estimated spectral distribution for the same three pairs of chosen grid points characterized by weak (a, d), mild (b, e) and strong tail dependence (c, f), respectively. The magenta bars show the estimates based on the 1950 samples in the test set. The estimates of evtGAN are very close to the ground truth for all cases, i.e., weak, mild and strong dependence (red lines in Fig.~\ref{fig:spectral}), except for some bias in the case of mild dependence in precipitation extremes (Fig.~\ref{fig:spectral}e). In contrast, the performance of the Brown--Resnick model (black lines in Fig.~\ref{fig:spectral}) is much more variable. It captures relatively well the two pairs with weak tail dependence (Fig.~\ref{fig:spectral}a, d) and does a decent job for strong dependence in temperature extremes (Fig.~\ref{fig:spectral}c) but fails completely for the remaining three cases (Fig.~\ref{fig:spectral}b,e, f). Furthermore, in contrast to Brown--Resnick, evtGAN is able to represent asymmetric dependence structures, as is evident from Fig.~\ref{fig:spectral}.

\begin{figure}[h]
    \centering
        \includegraphics[width=.91\textwidth]{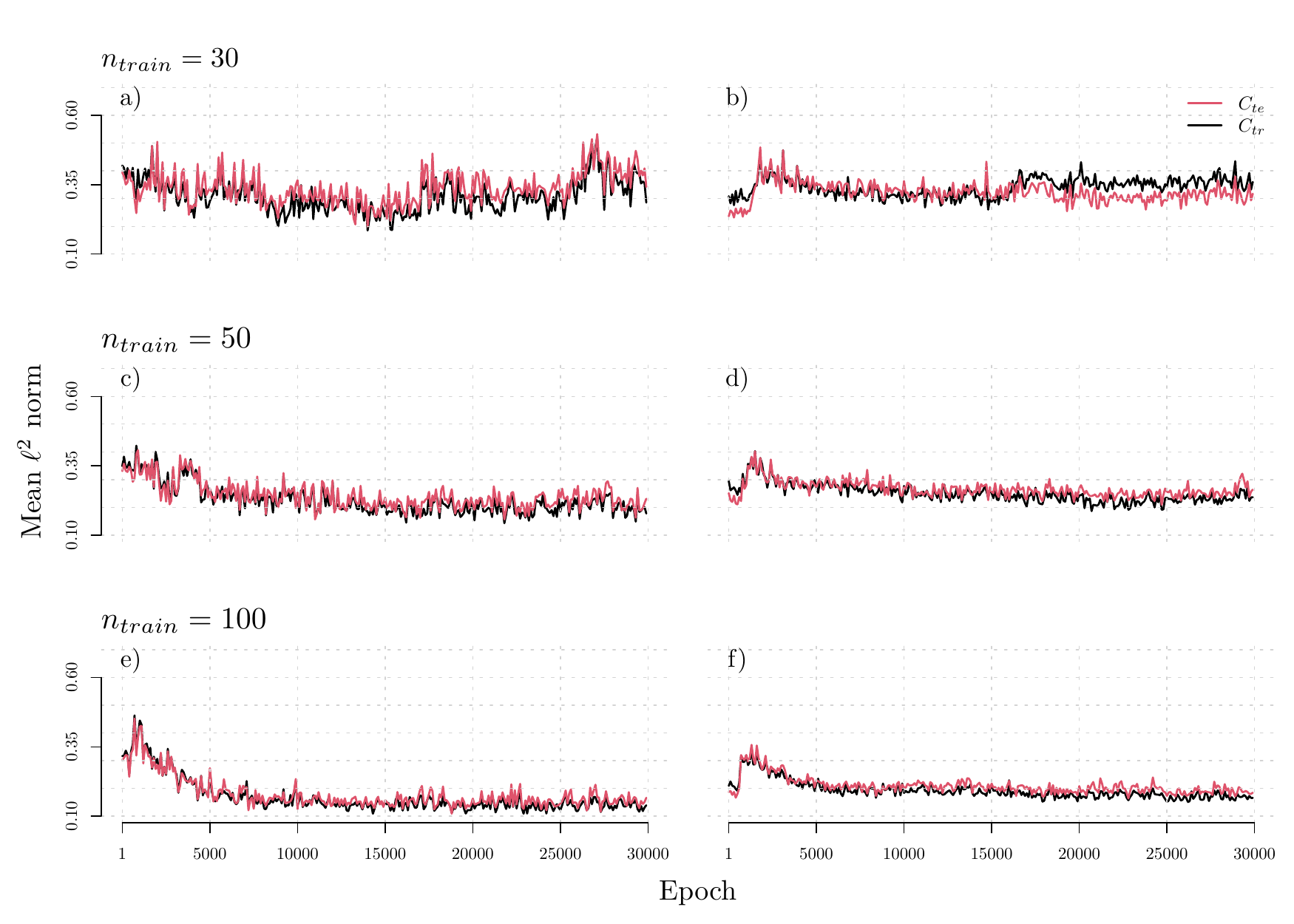}
        \caption{Model performance versus training epochs for different sample sizes in evtGAN for temperature (a, c, e) and precipitation (b, d, f) extremes. A number of observations equal to $n_{test}=N-n_{train}$ was sampled from evtGAN, where $N=2000$ for temperature, and $N=1999$ for precipitation. The mean $l^2$ norms for train (black) and test set (red) are defined as $C_{te}= ||\chi_{evtG}-\chi_{te}||_2$, $C_{tr}= ||\chi_{evtG}-\chi_{tr}||_2$, where $\chi_{evtG}$, $\chi_{te}$ and $\chi_{tr}$ denote the vectors of extremal correlations calculated on the samples from evtGAN, the test set and the train set, respectively}
    \label{fig:conv}
\end{figure}

We finally present a sensitivity analysis of the performance of evtGAN for different realistic sample sizes, namely 30, 50 and 100, while learning for 30,000 epochs (Fig.~\ref{fig:conv}). Overall the error tends to decrease the longer the models learn and it does not seem to matter whether we evaluate the performance on the train set or the test set (difference between black and red lines in Fig.~\ref{fig:conv} is small). For small sample sizes ($n=30$) the best performance might be reached for epochs smaller than 30,000 (Fig.~\ref{fig:conv}a, b) and the $l^2$ norm between the extremal coefficients of evtGAN and the train set could be used as a stopping criterion. An improvement in performance with increasing sample size is clearly visible (compare Fig.~\ref{fig:conv}a, b with Fig.~\ref{fig:conv}e, f). Furthermore, the model errors in temperature extremes are smaller than the ones for precipitation extremes (left versus right column in Fig.~\ref{fig:conv}). 
%Using this criterion as an early stopping method was initially considered, but the analysis quickly demonstrated, for $n_{train}=50$ and $n_{train}=100$, that using a fixed number of 30000 epochs provided slightly better results although the minimum is reached earlier. One could argue that the differences in terms of the criterion are not significant. However, for $n_{train}=30$, the criterion turned out to be useful, and could be used to approximate the number of epochs that should be considered, as it was sensitive to epoch training. For example, for temperature, epoch 13835 where the min is reached did yield better results than epoch 150000, which in turn produced better results than epoch 50000 and 100000. Interestingly, the differences in terms of the criterion for these epochs look as they would be significantly different. A heuristic that should still be rigorously verified but that would justify its use as far as are our results lead us to believe, is as long as the differences between the criterion computations at different epochs do not seem significantly different, use the latest one. If they seem significantly different, use the epoch that minimises the criterion. 

\section{Discussion}
\label{sec:dis}

The use of evtGAN combines the best of two worlds: correct extrapolation based on extreme value theory on the one hand, and flexible dependence modeling through GANs on the other hand.
GANs are an excellent tool to model complex dependencies in high-dimensional spaces. However, they are typically not tailored to model extremes in the marginals well. Indeed, for a standard DCGAN implementation where the marginals are not estimated by GEV distributions but empirically transformed, Figs.~\ref{fig:bivtemp}d, i, n and Figs.~\ref{fig:bivrain}d, i, n show that the generated samples are bounded by the range of the training sample. For an accurate extrapolation that resembles the marginal distributions of the test set, extreme value theory is required (panels c, e, h, j, m and o of Figs.~\ref{fig:bivtemp} and \ref{fig:bivrain}).

On the other hand, classical methods of spatial extreme value theory such as the Brown--Resnick process have accurate extrapolation properties for the marginal distributions. However, for an application to a spatially heterogeneous data set on a large domain (Fig.~\ref{fig:grid}) their parametric assumption may be too restrictive. Indeed, Figs.~\ref{fig:eval}c and f show a clear bias of the Brown--Resnick model in terms of bivariate extremal correlations, which is particularly visible for pairs with weak extremal dependence. Another indication for this bias can be seen in Figs.~\ref{fig:bivrain}j and o where the distributions of the Brown--Resnick samples differ strongly from the test set distributions. The evtGAN does not make prior assumptions on spatial stationarity or isotropy and therefore it does not exhibit a bias (points in Figs.~\ref{fig:eval}b and e are centered around the diagonal). This is particularly noteworthy since modeling complex non-stationarities for extremes is very difficult with parametric models \citep{hus2016,engelke2020graphical}. Considering the fitted bivariate distributions of evtGAN and Brown--Resnick underlines this point. 

The spectral distributions of the Brown--Resnick model are restricted to a parametric class of functions, which, for instance, are symmetric about $1/2$. The blue lines in Fig.~\ref{fig:spectral} show that this is too restrictive for our data since the strength of dependence is not correctly modelled (Fig.~\ref{fig:spectral}b) or the asymmetry is not captured (Fig.~\ref{fig:spectral}c,f). The evtGAN (red lines) on the other hand can model weak and strong dependences, and it even adapts to possible asymmetries. This is also evident from the scatterplots in Fig.~\ref{fig:bivtemp} and Fig.~\ref{fig:bivrain}, where the shape of the Brown--Resnick samples are restricted to a parametric sub-class of distributions, the so-called H\"usler--Reiss family \citep{hue1989}.

A further restriction of classical spatial max-stable processes is the fact that they are limited to modeling asymptotic dependence. For precipitation, it can be seen in Fig.~\ref{fig:eval}d-f that most of the test extremal correlations are close to zero, indicating asymptotic independence. While the evtGAN is able to capture this fairly well (Fig.~\ref{fig:eval}e), the Brown--Ressnick model always has positive extremal correlations, explaining the bias in the bottom left corner of Fig.~\ref{fig:eval}f. A spatial asymptotically independent model \citep[e.g.,][]{wadsworth2012dependence} would be a possible remedy for this, but such processes would still suffer from the limitation induced by non-stationarity and asymmetry described above.

Overall, evtGAN tends to perform better in capturing dependencies between temperature extremes than precipitation extremes (Fig.~\ref{fig:conv}). This is likely related to the fact that extremes in temperature are more spatially coherent \citep{Keggenhoff2014} (Fig.~\ref{fig:grid}).

\section{Conclusions}
\label{sec:conc}

Understanding and evaluating the risk associated with extreme events is of primal importance for society, as recently emphasized in the 6th Assessment Report of the Intergovernmental Panel on Climate Change \citep{Seneviratne2021}. Extreme event analysis and impact assessments are often limited by the available sample sizes. Furthermore, simulations with complex climate models are very expensive. Here we combine a machine learning approach with extreme value theory to model complex spatial dependencies between extreme events in temperature and precipitation across Europe based on a limited sample size. We demonstrate that this hybrid approach outperforms the typically used approach in multivariate extreme value theory and can well represent the marginal distributions and extremal dependencies across spatially distributed climate extremes. The approach can be easily adapted to other types of extremes and used to create large sample sizes that are often required for climate risk assessments.

%\paragraph{Acknowledgments}
%We are grateful for the technical assistance of A. Author.

\paragraph{Funding Statement}
This work was funded by the Swiss National Science Foundation (grant nos. 179876, 189908 and 186858) and the Helmholtz Initiative and Networking Fund (Young Investigator Group COMPOUNDX; grant agreement no. VH-NG-1537).

\paragraph{Competing Interests}
None

\paragraph{Data Availability Statement} The temperature and precipitation maxima alongside the scripts to reproduce the results in this study are available at https://doi.org/10.5281/zenodo.5821485.
%A statement about how to access data, code and other materials allowing users to understand, verify and replicate findings --- e.g. Replication data and code can be found in Harvard Dataverse: \verb+\url{https://doi.org/link}+.

%\paragraph{Ethical Standards}
%The research meets all ethical guidelines, including adherence to the legal requirements of the study country.

\paragraph{Author Contributions}
%Please provide an author contributions statement using the CRediT taxonomy roles as a guide {\verb+\url{https://www.casrai.org/credit.html}+}. 
Conceptualization: Y.B.; E.V.; S.E. Methodology: Y.B; E.V.; S.E. Formal analysis: Y.B. Software: Y.B.; E.V. Data curation: K.v.d.W. Data visualisation: Y.B. Supervision: J.Z.; S.E. Writing original draft: J.Z.; E.V.; S.E. All authors approved the final submitted draft.

\bibliographystyle{apalike}
%\bibliography{Sample-refs}

\bibliography{bibliography}

\end{document}